\documentclass{article}

\usepackage{microtype}
\usepackage{graphicx}
\usepackage{subcaption}
\usepackage{booktabs} 

\usepackage{hyperref}
\usepackage[ruled,vlined]{algorithm2e}

\usepackage[accepted]{icml2025}

\usepackage{amsmath}
\usepackage{amssymb}
\usepackage{mathtools}
\usepackage{amsthm}

\usepackage[capitalize,noabbrev]{cleveref}

\theoremstyle{plain}

\theoremstyle{definition}

\theoremstyle{remark}

\usepackage[textsize=tiny]{todonotes}

\icmltitlerunning{Submission and Formatting Instructions for ICML 2025}

\begin{document}

\twocolumn[
\icmltitle{TokenRing: An Efficient Parallelism Framework for Infinite-Context LLMs via Bidirectional Communication}



\icmlsetsymbol{equal}{*}

\begin{icmlauthorlist}
\icmlauthor{Zongwu Wang}{xxx, yyy}
\icmlauthor{Fangxin Liu*}{xxx, yyy}
\icmlauthor{Mingshuai Li\,$^2$}{}
\icmlauthor{Li Jiang*}{xxx, yyy}\\
\icmlauthor{1. Department of Computer Science and Engineering, Shanghai Jiao Tong University, \,\, 2. Shanghai Qi Zhi Institute}{}\\
\icmlauthor{*Corresponding Author \,\,\,\,\,\,\,\{wangzongwu,\,liufangxin,\,ljiang\_cs\}@sjtu.edu.cn}{}
\end{icmlauthorlist}



\icmlkeywords{Machine Learning, ICML}

\vskip 0.3in
]




\begin{abstract}
Efficient parallelization of Large Language Models (LLMs) with long sequences is essential but challenging due to their significant computational and memory demands, particularly stemming from communication bottlenecks in attention mechanisms. While sequence parallelism (SP) has been introduced as a potential solution, existing methods often suffer from limited scalability or inefficiency, rendering their effectiveness.

Ring-Attention demonstrates the potential for scaling sequence processing but faces significant limitations due to its reliance on peer-to-peer (P2P) communication and inefficient utilization of network resources. As the degree of SP increases, the quadratic decrease in computation time per step contrasts sharply with the linear reduction in communication volume, exacerbating communication bottlenecks. 
To address these challenges, we propose TokenRing, a fine-grained parallel framework that leverages bidirectional P2P communication to effectively overlap computation and data transmission. By partitioning the attention block and concurrently transmitting Query and block outputs (i.e., $block\_out$ and $block\_lse$) within a fully connected mesh topology, TokenRing achieves significant reductions in communication overhead and better load balancing. These innovations improve the scalability and efficiency of distributed Transformer models, particularly for long-context sequences. Experimental results demonstrate that TokenRing enhances throughput and reduces communication latency. Moreover, its design adapts seamlessly to various multi-GPU interconnect solutions, such as Huawei Ascend, ensuring broad compatibility and cost-effectiveness for distributed LLM inference and training. The code is available at: \url{https://github.com/ACA-Lab-SJTU/token-ring}.

\end{abstract}

\section{Introduction}
Large Language Models (LLMs), such as GPT \cite{brown2020language} and LLaMA \cite{touvron2023llama, touvron2023llama2, dubey2024llama} have emerged as a crucial part of artificial intelligence, showing outstanding performance in numerous applications like natural language processing, question answering, and code generation. As these models keep increasing in complexity and scale, there is a growing necessity for advanced parallelization techniques to enhance their training and inference efficiency. Among the key components of LLMs, the attention block \cite{vaswani2017attention} plays a vital role yet also brings about significant communication bottlenecks, especially when dealing with long-context sequences.

\begin{table*}[t]
\caption{Comparison of Different Parallelism}
\label{parallel-table}
\vskip 0.15in
\begin{center}
\begin{small}
\begin{sc}
\begin{tabular}{lll}
\toprule
Parallelism & Communication & limitation \\
\midrule
Tensor Parallelism~\cite{dean2012large} & AllReduce & memory in long context \\
Ring attention~\cite{liu2024blockwise} & single P2P SendRecv & communication bandwidth \\
DeepSpeed-Ulysses~\cite{jacobs2023deepspeed} & AlltoAll & number of attention heads \\
\textbf{TokenRing (Ours)} & \multicolumn{2}{l}{\textbf{\textcolor{blue}{bidirectional P2P SendRecv}}} \\
\bottomrule
\end{tabular}
\end{sc}
\end{small}
\end{center}
\vskip -0.1in
\end{table*}

Efficient parallelization methods for LLMs have been a major focus of research, with various strategies explored, each with its own strengths and limitations. Approaches such as data parallelism \cite{li2014scaling}, tensor parallelism \cite{dean2012large}, and pipeline parallelism \cite{huang2019gpipe} have been extensively studied. 
In data parallelism (DP), each worker maintains a complete copy of the model parameters, with training samples distributed uniquely across workers. Model computations are performed independently on each worker. However, as model size increases, DP introduces significant memory and communication overhead due to the need for parameter replication and synchronization during every iteration.
Pipeline parallelism (PP) divides the model into multiple stages distributed across GPUs to improve throughput. However, it can result in pipeline stalls and idle time if the workload distribution is uneven, decreasing overall efficiency.
In tensor parallelism (TP), individual operators of a model are partitioned across multiple workers. Each worker stores a portion of the operator's parameters and performs part of the computation, such as processing a tile of a matrix. While this approach reduces memory requirements per worker, it introduces complex communication patterns and incurs additional memory overhead for tensor redistribution. Sequence Parallelism (SP), a technique that involves partitioning a single input sequence, has emerged as a particularly promising approach for training or inference tasks involving long-context sequences. Notable methods such as Ring Attention \cite{liu2024blockwise, liu2024ringattention} and DeepSpeed-Ulysses \cite{jacobs2023deepspeed} have demonstrated significant potential in this area. However, several challenges persist. For instance, the parallelism degree of SP in DeepSpeed-Ulysses is constrained to a value lower than the number of attention heads, while communication efficiency in Ring Attention suffers from degraded performance due to suboptimal point-to-point (P2P) bandwidth utilization. These limitations currently impede the broader adoption of Sequence Parallelism in distributed Transformer computations.

In this work, we present \textbf{TokenRing}, a communication-oriented parallelism framework designed to address the communication overhead introduced by partitioning the attention block across multiple GPUs to accommodate increased token lengths. TokenRing achieves scalability by distributing the QKV tensors along the token dimension across GPUs and integrating SP.
To further optimize communication efficiency in attention blocks, TokenRing leverages inter-node P2P bidirectional communication bandwidth (in Table \ref{parallel-table}). Unlike traditional methods that simultaneously send and receive KV chunks for the next micro-step, TokenRing enables the concurrent transmission of both the query (Q) and block outputs (block\_out and block\_lse). This reduces idle communication time and provides more opportunities for computation-communication overlap. As a result, TokenRing accelerates processing and supports the integration of long-context LLMs in resource-constrained environments.
Our results demonstrate that TokenRing significantly reduces communication time and increases throughput, leading to substantial performance improvements. Furthermore, it utilizes widely available NVLink connections in Nvidia GPU architectures and adapts to other multi-GPU interconnect solutions, such as Huawei Ascend. This flexibility ensures compatibility with diverse hardware configurations, making TokenRing a practical and cost-effective solution for distributed LLM inference and training.

In summary, we make the following contributions:
\begin{itemize}
    \item \textbf{Communication-Efficient Parallelism Design.} We propose a novel approach that partitions QKV tensors along the token dimension across GPUs, enabling efficient handling of long token sequences. TokenRing leverages inter-node P2P bidirectional communication to reduce communication overhead and enhance scalability.
    \item \textbf{Optimization of Attention Block Communication.} 
    We introduce a mechanism that utilizes bidirectional bandwidth to concurrently transmit the query (Q) and block outputs ($block\_out$ and $block\_lse$), as opposed to previous solutions that relied on unidirectional bandwidth. This approach enhances communication efficiency by fully utilizing the available bandwidth, improving computation-communication overlap, and accelerating processing in attention mechanisms.

    \item \textbf{Flexible and Adaptable Design. }TokenRing is compatible with various models, including DIT and LLMs, and can integrate with other parallelization methods. It efficiently utilizes widely available NVLink connections in Nvidia GPU architectures and adapts to other multi-GPU interconnect solutions, such as Huawei Ascend. This flexibility allows for the selection of suitable configurations tailored to different application scenarios while reducing dependency on high-cost solutions like NVSwitch.
\end{itemize}

\section{Background}

\subsection{Sequence Parallelism}
\begin{figure}[t]
\centering
\begin{subfigure}{0.5\textwidth}
\includegraphics[width=\textwidth]{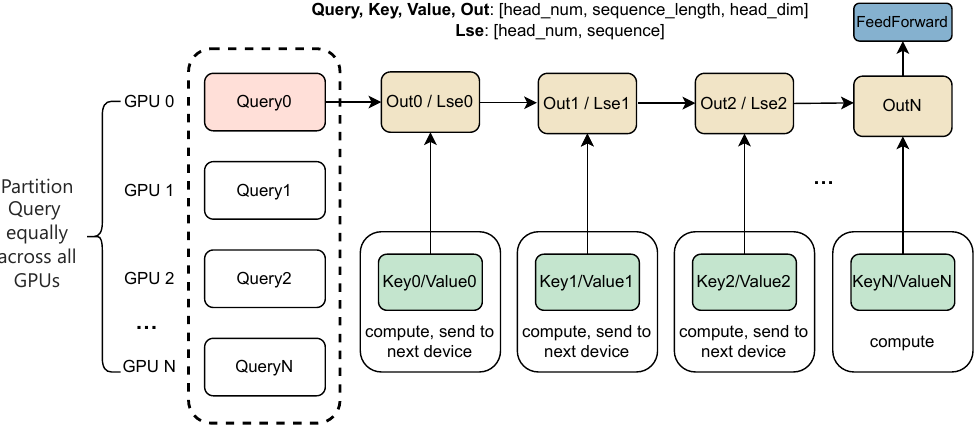}
\caption{Ring-Attention Mechanism.}
\label{subfig:ringattention}
\end{subfigure}
\quad
\begin{subfigure}{0.5\textwidth}
\includegraphics[width=\textwidth]{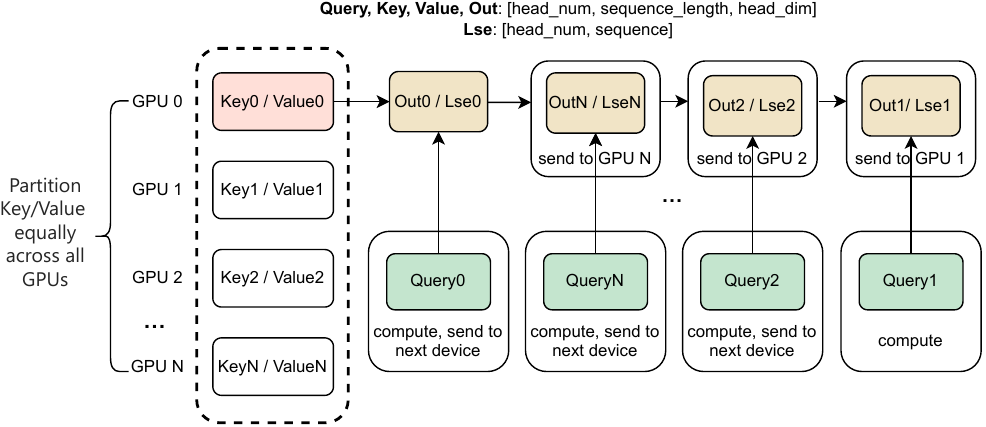}
\caption{TokenRing Mechanism.}
\label{subfig:tokenring}
\end{subfigure}
\caption{Comparison of the Ring-Attention (a) and TokenRing (b) overviews. In TokenRing, each GPU stores a single key-value block, while query blocks circulate through the ring for processing. The process begins with an initial query block, which is iterated over along with other query blocks. These query blocks, in conjunction with the key-value blocks, are utilized to compute self-attention using flash attention. Simultaneously, the ($block\_out$) and ($block\_lse$) are transmitted to the corresponding GPUs to update the outputs in reverse order.}
\label{fig:overview}
\end{figure}

SP has become essential for enhancing the training and inference efficiency of LLMs, particularly when managing long sequences. Among the various SP approaches, Ring Attention stands out for its ability to address the memory constraints of transformers handling extended sequences. It achieves this by partitioning self-attention and feedforward computations into blocks distributed across multiple devices arranged in a ring topology. This arrangement allows for simultaneous computations of key-value (KV) blocks and communication. Figure~\ref{subfig:ringattention} shows the Ring attention mechanism.

However, Ring-Attention faces significant communication inefficiencies, primarily due to the communication latency introduced by the P2P transmission of KV chunks over the network, which becomes a major bottleneck. Specifically, as Ring-Attention scales, the computation time per step decreases quadratically, while the communication volume per step decreases only linearly. This disparity exacerbates the imbalance between computation and communication. Additionally, the ring-based communication design leads to suboptimal utilization of network resources, as it wastes bandwidth in one direction, preventing the full utilization of available bandwidth and missing opportunities to further reduce communication overhead and optimize performance.

In contrast, Ulysses maintains a constant communication volume by proportionally increasing both the sequence length and the number of devices. It partitions the Q, K, V, and output tensors along the sequence dimension and employs All2All communication to shift partitioning to the attention head dimension. This allows for more efficient attention computation for each head. However, Ulysses is limited by its degree of parallelism, which cannot exceed the number of attention heads. This restriction limits its applicability in scenarios involving Grouped Query Attention (GQA) \cite{ainslie2023gqa} or Multi-Query Attention (MQA) \cite{shazeer2019fast}.

Megatron-LM optimizes the AllReduce operation within Tensor Parallelism by replacing it with equivalent Allgather and Reduce-Scatter operations on partitioned data, similar to ZeRO-2 \cite{ren2021zero}. This adjustment reduces the memory overhead associated with activations while maintaining comparable communication costs. However, it is important to note that Megatron-LM's Sequence Parallelism is not standalone; it must be integrated with tensor parallelism to be effective.

\subsection{Interconnection Preliminary}
In the context of LLMs, the connectivity modalities and topological architectures of GPUs are of paramount importance. Peripheral Component Interconnect Express (PCIe) is widely employed due to its broad compatibility; however, it offers limited bandwidth for GPU-specific communications. In contrast, NVIDIA's NVLink is tailored to meet the high-bandwidth demands of GPU-to-GPU data transfers, thereby facilitating parallel processing and enhancing the efficiency of model training and inference. Furthermore, NVSwitch, another innovation from NVIDIA, serves as a key switching fabric that interconnects multiple GPUs, providing enhanced connectivity and flexibility for effective data distribution within a GPU cluster. Other manufacturers also offer alternative interconnection solutions; for instance, Huawei's Collective Communication Server (HCCS) provides high-speed direct connections between chips, thus bolstering communication capabilities within a GPU cluster.

\begin{figure}[t]
\centering
\includegraphics[width=0.4\textwidth]{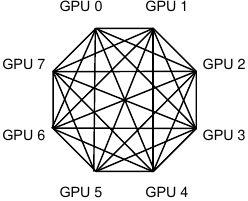}
\caption{Topology of OCP Accelerator Module.}
\label{fig:OAMtopo}
\end{figure}

When discussing topological configurations, two primary paradigms dominate the field. The OCP Accelerator Module (OAM) architecture, commonly adopted by manufacturers other than NVIDIA, interconnects GPUs within a node using a combination of interfaces such as PCIe and HCCS, as illustrated in Figure \ref{fig:OAMtopo}. This architecture is particularly well-suited for all-to-all operations, which involve simultaneous data transfers among all processing elements. However, the OAM topology often employs a full mesh interconnect model. In a configuration with 8 GPUs interconnected in a full mesh OAM topology, the direct bandwidth available between any two GPUs is approximately $1/8$ of the total aggregate bandwidth. This reduction can impede communication efficiency within a single machine, especially when managing large data transfers and complex computational tasks that require rapid and widespread data sharing among GPUs.

\begin{figure}[t]
\centering
\includegraphics[width=0.5\textwidth]{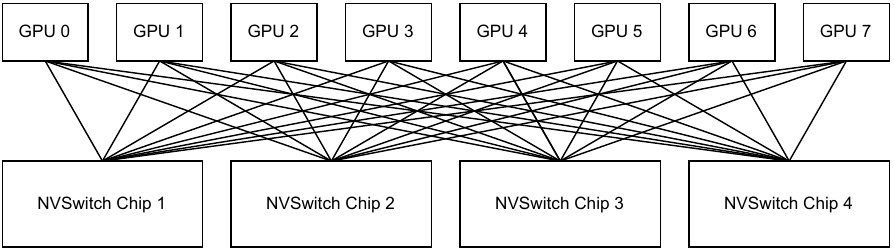}
\caption{Topology of the NVIDIA NVLink Switch System.}
\label{fig:nvswitch}
\end{figure}

In contrast, the NVSwitch-based topology developed by NVIDIA connects all GPUs to NVSwitch via NVLink, as depicted in Figure \ref{fig:nvswitch}, thus providing higher P2P communication bandwidth. Despite its advantages, the implementation cost of NVSwitch can be substantial, making it less viable for projects with budget constraints. Moreover, when multiple GPUs attempt to communicate simultaneously, NVSwitch may experience congestion and contention issues. As the number of concurrent communication requests increases, the switch may struggle to effectively manage data flow, leading to bottlenecks and suboptimal utilization of available bandwidth. This can lead to performance degradation and longer processing times, especially in applications that require seamless, concurrent communication across a large number of GPUs. Therefore, designing a GPU cluster for LLM computation requires a careful balance between various interconnection interfaces and topological architectures to optimize both performance and cost-efficiency.


\subsection{Recent works on Long-Context LLM}

In large-scale clusters, there are primarily two approaches for handling long-context LLM inference. The first is chunked prefill \cite{agrawal2023sarathi}, which divides prefill requests into equal-sized chunks and applies attention masks to ensure computational equivalence. The second approach involves decoupling the prefill and decoding phases \cite{zhong2024distserve, qin2024mooncake}, thoroughly separating both computational operations and hardware resources. The prefill phase closely resembles the forward phase in training, allowing us to reference relevant training research. 

USP \cite{fang2024unified} ingeniously combines DeepSpeed's Ulysses and Ring Attention. By employing a hybrid parallel architecture, it eliminates the head-number constraint present in SP-Ulysses and implements a robust communication schema suitable for heterogeneous networks. LoongTrain \cite{gu2024loongtrain, chen2024internevo} introduces the 2D-Attention and Double-Ring-Attention mechanisms, which merge head-parallel and context-parallel approaches, thereby overcoming scalability limitations and enhancing efficiency. DistFlashAttn \cite{li2021sequence} adopts a new load-balancing and scheduling technique to overlap computation and communication in causal language models. Tree Attention \cite{shyam2024tree} proposes an energy function for self-attention, utilizing gradient computation to optimize communication among multiple GPUs based on a tree topology. In the context of multi-turn dialogue, the prefill stage can be further categorized into full prefill and partial prefill \cite{yang2024context}, allowing for the selection of more appropriate parallel strategies accordingly.

In existing distributed systems, many approaches employ hybrid parallelization strategies to address challenges such as head-number constraints and scalability limitations in self-attention mechanisms. However, these methods often do not effectively leverage bidirectional communication. Additionally, lightweight solutions that consider intra-node hardware topology and maximize inter-GPU bandwidth utilization remain underexplored, potentially hindering further advancements in performance and efficiency.

\section{TokenRing Sequence Parallelism}
We introduce TokenRing, a scalable and efficient framework designed to tackle the challenges of executing LLMs with long sequences.
TokenRing is built on distributed attention, which integrates sequence parallelism into the attention mechanism. This approach effectively addresses the scalability limitations of traditional hybrid parallelism methods, such as head-parallel attention combined with context-parallel attention. Additionally, to minimize communication overhead within attention blocks, we design a lightweight ring mechanism that leverages bidirectional communication within a node, further enhancing efficiency.

\subsection{Motivation and Key Insights}
Handling long-context scenarios in existing Ring Attention methodologies presents significant challenges, particularly under constrained GPU memory scenarios. Communication overhead emerges as a critical bottleneck, especially as the number of GPUs increases~\cite{gu2024loongtrain}. While computation time per step decreases quadratically, the communication volume per step reduces only linearly, exacerbating this bottleneck. Additionally, when GPUs are interconnected via NVLink or similar OAM architectures, the original Ring Attention mechanism utilizes only a unidirectional portion of the ring bandwidth, leaving much of the available capacity underutilized.

In typical self-attention mechanisms, the sequence length $N$, number of attention heads $H$, and head dimension $D$ define the dimensions of the query ($Q$), key ($K$), and value ($V$) tensors as ($N$,$H$,$D$). A promising approach to alleviate communication overhead involves transmitting $Q$ tensors instead of $KV$ pairs, as $Q$ is smaller in size for large $N$. However, this strategy introduces additional data transmission requirements, including block-wise outputs ($block\_out$) and log-sum-exp ($block\_lse$) values, which are computed at each step using flash attention mechanisms\cite{dao2022flashattention, dao2023flashattention2}. These updates are mathematically defined as:
\begin{align*}
out &= out - \sigma(block\_lse - lse) \times (out - block\_out)\\
lse &= lse - \ln(\sigma(lse - block\_lse))
\end{align*}
where $\sigma$ represents the sigmoid function.  This formulation captures the dynamics of updating $lse$ and $out$ within the TokenRing framework, ensuring accurate propagation and refinement of intermediate computations during the attention process. However, the dimensions of $block\_out$ ($N$,$H$,$D$) and $block\_lse$($H$,$N$) result in a substantial communication volume, thereby offsetting the expected benefits of transmitting smaller tensors like $Q$.

By enabling concurrent bidirectional transmission of $Q$ and $Out$ tensors to different GPUs, TokenRing maximizes the use of available interconnect bandwidth. This strategy significantly reduces communication time, enhancing the efficiency of processing long-context sequences within the constraints of current hardware architectures.

\subsection{TokenRing Parallelism Framework}
\begin{algorithm}[tb]
   \caption{TokenRing Parallelism}
   \label{alg:token_ring}
\begin{algorithmic}
   \STATE {\bfseries Input:} Q, K, V,    Number of hosts $N$
   \STATE 
   Split Q, K, V into N blocks
   On GPU $j \in [0, 1, ..., N-1]$
   \FOR{$i=0$ {\bfseries to} $N-1$}
  \STATE s $\leftarrow (j + 1) \mod N $
   \IF{$i < N - 1$}
   \STATE Rank $j$ async\_send $Q_j^i$ to rank $s$
   \ENDIF
    \STATE $t$ $\leftarrow (j - step + 1) \mod N $
   \IF{$i > 1$}
   \STATE Rank $j$ async\_send $O_j^i$, $L_j^i$ to rank $t$
   \ENDIF

   \STATE $block\_out$, $block\_lse$ $\leftarrow Attention(Q_j^i, K_j, V_j) $
   \STATE synchronize()

  \IF{$i \: != 1$}
   \STATE $Out$, $Lse$ $\leftarrow Update$($block\_out$,  $ block\_lse$) 
   \ENDIF
   
   \ENDFOR
   \STATE send $block\_out$, $block\_lse$ to rank j - N + 1
   \STATE synchronize()
   \STATE $Out$, $Lse$ $\leftarrow Update$($block\_out$,  $ block\_lse$) 
\end{algorithmic}
\end{algorithm}

In this subsection, we present the \textbf{TokenRing} parallelism framework, which addresses communication inefficiencies in multi-GPU environments for LLMs. The pseudocode is provided in Algorithm~\ref{alg:token_ring}.

TokenRing first partitions the input tensors \(Q\), \(K\), and \(V\) into \(N\) blocks, where \(N\) is the number of GPUs. Each GPU \(j\) executes an iterative compute blockwise attention to balance communication and computation.

The overview of TokenRing is shown in Figure~\ref{subfig:tokenring}. During iteration \(i\), for \(i < N - 1\), GPU \(j\) asynchronously transmits the \(i\)-th block of \(Q_j\) to the next GPU in the ring, denoted as rank \(s = (j + 1) \bmod N\). Concurrently, for \(i > 1\), GPU \(j\) sends the intermediate outputs \(O_j^i\) and log-sum-exp values \(L_j^i\) to another GPU in the ring, rank \(t = (j - \text{step} + 1) \bmod N\). This bidirectional communication scheme is designed to overlap with computation, minimizing idle time and maximizing throughput. 
\begin{figure}[t]
\centering
\includegraphics[width=0.5\textwidth]{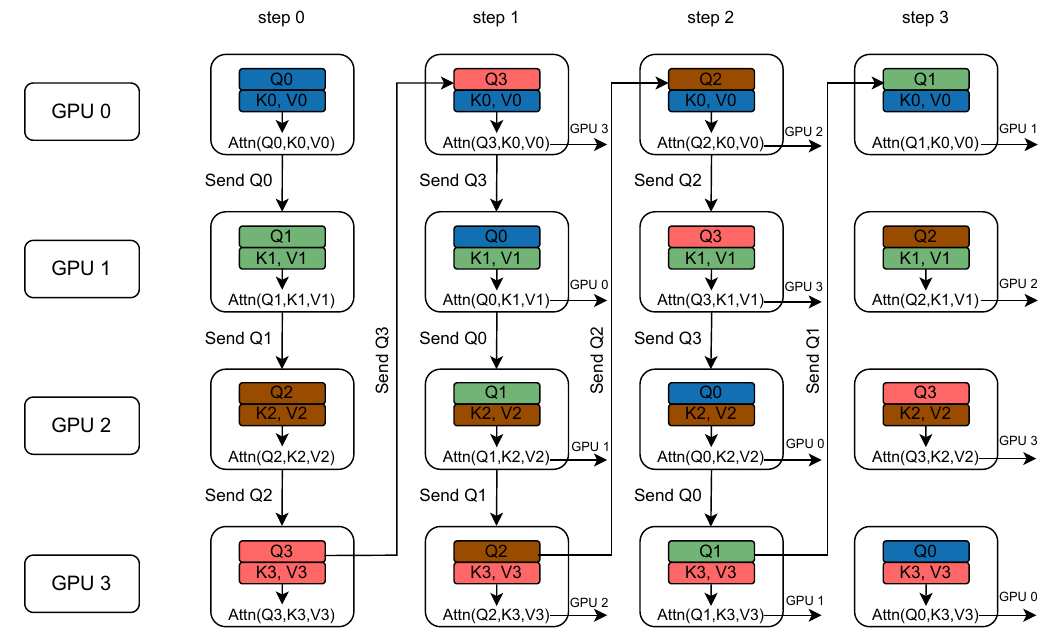}
\caption{Implementation of the TokenRing in xDIT Framework Utilizing Four GPUs.}
\label{fig:tokenring}
\end{figure}

It is worth noting that our method is orthogonal to Flash Attention, enabling the utilization of its efficient computation features. During computation, Flash Attention is utilized to compute \(block\_out\) and \(block\_lse\) on each GPU using \(Attention(Q_j^i, K_j, V_j)\). After computation, a synchronization step ensures data consistency across GPUs. For \(i \neq 1\), an update step refines the \(Out\) and \(Lse\) values based on the computed \(block\_out\) and \(block\_lse\).


Once all iterations are complete, the final \(block\_out\) and \(block\_lse\) are sent to a specific GPU in the ring, rank \(j - N + 1\), followed by a final synchronization step to ensure consistency. The final output \(Out\) and \(Lse\) are then derived using the \(Update\) function, integrating information from all blocks.

The TokenRing introduces a novel communication-computation orchestration leveraging the ring topology. By simultaneously transmitting \(Q\) and \(block\_out\) in opposite directions, it reduces communication latency, maximizes bandwidth utilization, and significantly enhances the efficiency of Transformer model training and inference.

\subsection{Application and Implementation}
\subsubsection{Case Study I: Diffusion Transformers} 
To demonstrate the implementation of the TokenRing algorithm, we utilize xDIT \cite{sun2024unveiling, fang2024xdit}, an inference engine designed for large-scale model parallelism. Notably, xDIT natively supports SP, aligning seamlessly with the data partitioning requirements of TokenRing. The input tensors \(Q\), \(K\), and \(V\) are initially partitioned into \(N\) blocks based on the number of GPUs. Each GPU then executes iterative computations as defined by the TokenRing algorithm. Figure~\ref{fig:tokenring} depicts the TokenRing architecture deployed on four GPUs within xDIT, highlighting core components such as data partitioning, bidirectional communication of \(Q\) and \(block\_out\), and synchronization steps to maintain consistency across GPUs. Figure~\ref{fig:singleGPU} shows in more detail all the tasks executed on a single GPU.

\textbf{Step 0:} 
In the initial step, each GPU performs both data transmission and computation simultaneously. Specifically, each GPU transmits its segment of \(Q\) to the GPU with rank \(+1\) while simultaneously receiving new \(Q\) data from the GPU with rank \(-1\). Currently, Flash Attention 2 computes the attention outputs for the current block of \(Q\), \(K\), and \(V\), generating \(block\_out\) and \(block\_lse\). A key feature of our approach is the utilization of bidirectional bandwidth for communication, which reduces communication overhead and prevents communication from becoming a bottleneck. This enables better parallelism and reduces idle times, allowing for more efficient processing. Once both operations are completed, the \(Q\) and \(Out\) are updated, ensuring data freshness for subsequent steps.

\textbf{Step 1:} In this step, each GPU utilizes the newly received \(Q\) from the previous step to perform fresh attention computations. Simultaneously, it forwards its own \(Q\) segment to the next GPU. As \(block\_out\) and \(Out\) are distributed across GPUs, no immediate \(Out\) update is required during this step. This phase sustains the momentum of data transmission and computation across the system.

\textbf{Step 2:} Alongside the usual \(Q\) transmission and attention computation, a critical step occurs: \(block\_out\) and \(block\_lse\) from the previous step are sent to the preceding GPU (rank \(-1\)). This enables the \(Out\) update across GPUs, aggregating and refining intermediate results. This step exemplifies the algorithm's efficiency, leveraging bidirectional data flow to optimize both communication and computation.

\textbf{Step 3:} In the final step, the transmission of \(Q\) stops. Instead, \(block\_out\) from Step 2 is relayed to the GPU with rank \(-2\) while attention computations proceed as normal. After this step, an additional communication phase occurs, transmitting \(block\_out\) to the GPU with rank \(-3\) for the final \(Out\) update. This concludes the computation, yielding the ultimate \(Out\).

It is worth noting that in a four-GPU setup, the benefits of TokenRing may appear modest, as only one step fully exploits bidirectional communication, and an additional communication phase is required at the end. However, as the number of GPUs increases, the proportion of steps utilizing bidirectional communication grows, significantly reducing communication latency and boosting overall computational efficiency. Thus, the TokenRing algorithm emerges as a scalable and efficient solution for Diffusion Transformers \cite{peebles2023scalable} in multi-GPU environments, well-suited to meet the computational demands of complex generative tasks.

\subsubsection{Case Study II: Large Language Models} 

In the inference of large language models (LLMs), a distinct challenge arises compared to architectures like DiT: the inherent causality of the attention mechanism. Specifically, during attention computation, each query (\(Q\)) tensor interacts only with the key (\(K\)) and value (\(V\)) tensors corresponding to its current and preceding positions in the sequence. This causal structure leads to an uneven distribution of computational loads when tensors are naively partitioned along the sequence length.

To address this imbalance and improve efficiency, two partitioning strategies---the striped strategy \cite{brandon2023striped} and the zigzag strategy \cite{ring-flash-attention}---have been proposed. In this work, we adopt the zigzag strategy, which integrates seamlessly with TokenRing to minimize communication overhead and reduce data transfer volume. For instance, in a four-GPU setup where the sequence length is divided into eight segments, GPU 0 processes the 0th and 7th segments in step 0. After completing computation for the 0th segment, the corresponding \(Q\) tensor is no longer needed in subsequent steps, eliminating redundant data transmission. This pattern of releasing unused data continues throughout the inference process, significantly reducing communication cost.

By dynamically aligning data transmission with the causal computation pattern, the combination of the zigzag strategy and TokenRing achieves balanced workloads and efficient resource utilization. The bidirectional data flow inherent to TokenRing further reduces communication latency by overlapping tensor transfers with computation. This synergy ensures that the challenges of uneven workloads and excessive communication in LLM inference are effectively mitigated, delivering improved performance for large models with extended sequence lengths.

\subsubsection{Case Study III: Multi-Node Distributed System}

\begin{figure}[t]
\vspace{-0.1cm}
\setlength{\abovecaptionskip}{3pt} 
\setlength{\belowcaptionskip}{3pt}
\centering
\includegraphics[width=0.93\linewidth]{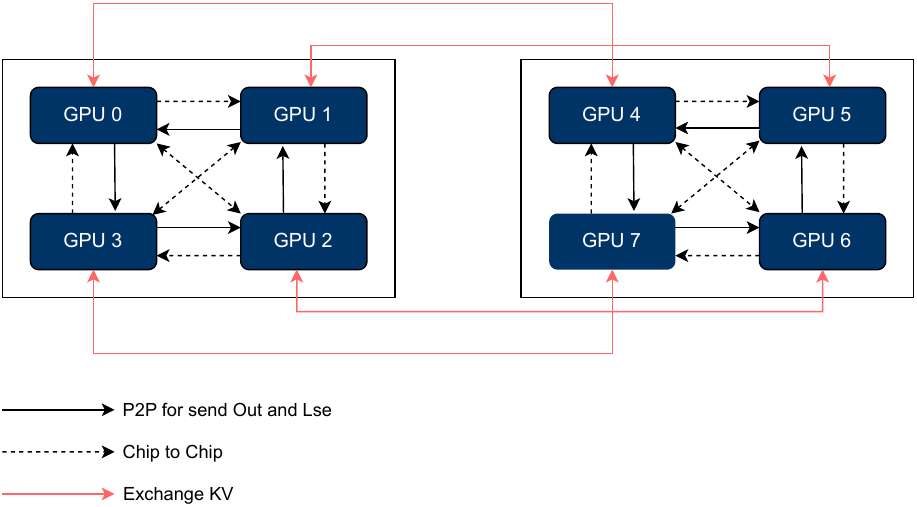}
\caption{TokenRing Implementation Across Two Nodes.}
\label{fig:amongnode}
\vspace{-0.5cm}
\end{figure}

For multi-node scenarios, the standalone use of TokenRing faces inherent limitations. Each step must wait for cross-node communication to finish, and the lack of full interconnection between GPUs across nodes restricts the full utilization of TokenRing's advantages. To overcome these constraints, we propose a hybrid approach that combines TokenRing with the original Ring Attention mechanism, as shown in Figure~\ref{fig:amongnode}. 

In such a hybrid scheme, computations are first performed independently within each node using TokenRing. During this process, \(block\_out\) and \(block\_lse\) are transmitted locally. Once partial computations are completed, the corresponding \(K\) and \(V\) tensors on GPUs across nodes are exchanged to facilitate the remaining computations. Specifically, Ring Attention is employed for cross-node communication of \(K\) and \(V\), while TokenRing is utilized within individual nodes for intra-node communication and computation. 

This multi-node extension effectively distributes the computational load and ensures the seamless exchange of necessary data. By leveraging both inter- and intra-node communication, the hybrid approach significantly enhances the scalability and efficiency of the system. The exchange of \(K\) and \(V\) tensors between nodes ensures a smooth continuation of the computation process, allowing the combined computational power of multiple nodes to be fully utilized. 

TokenRing enables the system to handle large-scale data processing and computationally intensive tasks. By accelerating complex algorithm execution and improving overall performance and throughput, this multi-node extension of TokenRing provides a robust and scalable solution for next-generation distributed computing environments.


\section{Performance Evaluation}
\subsection{Experiment Setup}
To evaluate the performance of the TokenRing algorithm, we conduct both unit tests and integrate TokenRing into the XDIT framework. Given the requirement for a full-mesh topology, testing on hardware optimally configured for TokenRing is not feasible for us. As a result, the experiments were performed on NVIDIA A10 GPUs. Within the test setup, four GPUs were housed in the same node and interconnected via PIX (connections traversing at most one PCIe bridge) and PXB (connections traversing multiple PCIe bridges without crossing the PCIe Host Bridge). 
The inference performance of TokenRing was evaluated using the LLaMA2-7B model configuration, with \(d = 128\) and \(nheads = 32\) for the multi-head attention (MHA) mechanism. This setup allowed us to examine the algorithm's efficiency under realistic hardware constraints and measure its potential for practical deployment in large-scale inference tasks.

\subsection{EVALUATION RESULTS}

\begin{figure}[t]
\centering
\begin{subfigure}{0.5\textwidth}
\includegraphics[width=\textwidth]{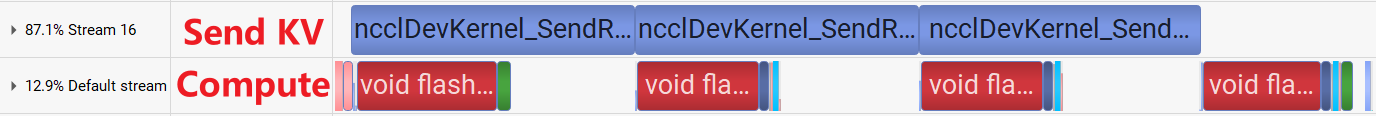}
\caption{Ring attention}
\end{subfigure}
\quad
\begin{subfigure}{0.5\textwidth}
\includegraphics[width=\textwidth]{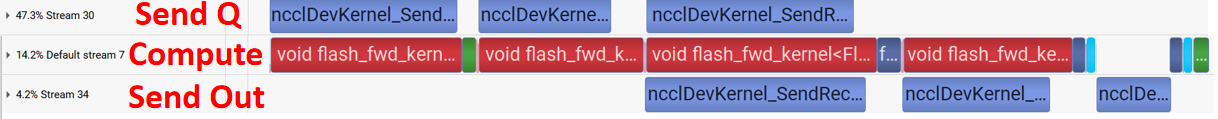}
\caption{Token ring}
\end{subfigure}
\caption{Inference profiling of the attention mechanism with a sequence length of 24,000.}
\label{fig:nsys}
\end{figure}
We present results for attention with a sequence length of 24,000 on a 4-card A10 platform, as shown in Figure~\ref{fig:nsys}. The results are obtained using the Nsight Systems profiler. The plot shows that under the sequence length of 24,000, TokenRing overlaps \(Q\) transmission and computation effectively during Step 0 and Step 1, achieving a time of approximately 3.5 ms. In Step 2, where \(Q\) and \(Out\) are transmitted simultaneously with ongoing computation, the time slightly increases to around 4.6 ms due to the PXB connection mode.
In comparison, Ring Attention suffers significantly longer communication times than computation, making communication the bottleneck in each round with a time of about 7.6 ms. This highlights the advantage of TokenRing’s bidirectional communication, which becomes more pronounced as the number of GPUs increases and the GPU connection topology transitions to a full-mesh structure.


\section{Discussion and Conclusion}
In this work, we introduce TokenRing, a fine-grained parallel framework designed to optimize communication efficiency in multi-GPU environments, particularly for processing long-context sequences in Transformer models. TokenRing addresses a critical challenge in distributed systems---such as in Ring Attention---where communication and computation cannot be effectively overlapped. By leveraging bidirectional communication, TokenRing enables the concurrent transmission of Query and block-out tensors, significantly reducing communication overhead compared to traditional methods.

We further present a detailed implementation of TokenRing, demonstrating how it reduces communication time and increases throughput. By integrating TokenRing into the xDIT framework and applying it to large language models using the zigzag strategy, we show significant improvements in both communication efficiency and computational load balancing. This framework is highly adaptable and can be applied to a wide range of models, ensuring scalability in multi-GPU setups.

However, several challenges were encountered during implementation. Despite the use of PyTorch's NCCL backend, which facilitates data transmission through two independent channels, GPU resource preemption has occasionally led to increased latency. This has prevented the full realization of the theoretical performance gains. Future work should focus on better integrating the communication strategy with the underlying hardware architecture. Tailoring the approach to the specific capabilities of different hardware platforms will further optimize TokenRing's performance. Such advancements will unlock its full potential, enabling faster training and inference for large models with long-context sequences and improving the overall efficiency of distributed computing systems.

\section*{Acknowledgements}
This work is sponsored by the National Natural Science Foundation of China (No.62402311) and CAAI-Ant Group Research Fund.



\newpage





\bibliography{example_paper}
\bibliographystyle{icml2025}

\newpage
\appendix
\onecolumn
\section{You \emph{can} have an appendix here.}

You can have as much text here as you want. The main body must be at most $8$ pages long.
For the final version, one more page can be added.
If you want, you can use an appendix like this one.  

The $\mathtt{\backslash onecolumn}$ command above can be kept in place if you prefer a one-column appendix, or can be removed if you prefer a two-column appendix.  Apart from this possible change, the style (font size, spacing, margins, page numbering, etc.) should be kept the same as the main body.

\end{document}